\documentclass[sigconf]{acmart}

\usepackage{subfloat}
\usepackage{subfig}
\usepackage{multicol}
\usepackage{tabularx}
\usepackage{amsmath,amstext}
\usepackage{needspace}
\usepackage{url}
\usepackage{enumerate}
\usepackage{algorithm,algpseudocode}
\usepackage{tikz}

\AtBeginDocument{%
  }

\copyrightyear{2022}
\acmYear{2022}
\setcopyright{acmlicensed}\acmConference[ICMI '22]{INTERNATIONAL
CONFERENCE ON MULTIMODAL INTERACTION}{November 7--11,
2022}{Bengaluru, India}
\acmBooktitle{INTERNATIONAL CONFERENCE ON MULTIMODAL INTERACTION
(ICMI '22), November 7--11, 2022, Bengaluru, India}
\acmPrice{15.00}
\acmDOI{10.1145/3536221.3556584}
\acmISBN{978-1-4503-9390-4/22/11}

\begin{document}

\title{Supervised Contrastive Learning  for Affect Modelling}

\author{Kosmas Pinitas}
\affiliation{%
  \institution{Institute of Digital Games, University of Malta}
  \city{Msida}
  \country{Malta}}
\email{kosmas.pinitas@um.edu.mt}

\author{Konstantinos Makantasis}
\affiliation{%
  \institution{Institute of Digital Games, University of Malta}
  \city{Msida}
  \country{Malta}}
\email{konstantinos.makantasis@um.edu.mt}

\author{Antonios Liapis}
\affiliation{%
  \institution{Institute of Digital Games, University of Malta}
  \city{Msida}
  \country{Malta}}
\email{antonios.liapis@um.edu.mt}

\author{Georgios N. Yannakakis}
\affiliation{%
  \institution{Institute of Digital Games, University of Malta}
  \city{Msida}
  \country{Malta}}
\email{georgios.yannakakis@um.edu.mt}

\renewcommand{\shortauthors}{Pinitas et al.}

\begin{abstract}

  
  
  

  Affect modeling is viewed, traditionally, as the process of mapping measurable affect manifestations from multiple modalities of user input to affect labels. That mapping is usually inferred through end-to-end (manifestation-to-affect) machine learning processes. What if, instead, one trains general, subject-invariant representations that consider affect information and then uses such representations to model affect?  
  In this paper we assume that affect labels form an integral part, and not just the training signal, of an affect representation and we explore how the recent paradigm of contrastive learning can be employed to discover general high-level affect-infused representations for the purpose of modeling affect. We introduce three different supervised contrastive learning approaches for training representations that consider affect information. In this initial study we test the proposed methods for arousal prediction in the RECOLA dataset based on user information from multiple modalities. 
  Results demonstrate the representation capacity of contrastive learning and its efficiency in boosting the accuracy of affect models. Beyond their evidenced higher performance compared to end-to-end arousal classification, the resulting representations are general-purpose and subject-agnostic, as training is guided though general affect information available in any multimodal corpus. 
  
\end{abstract}

\begin{teaserfigure}
\centering
\includegraphics[width=0.71\linewidth]{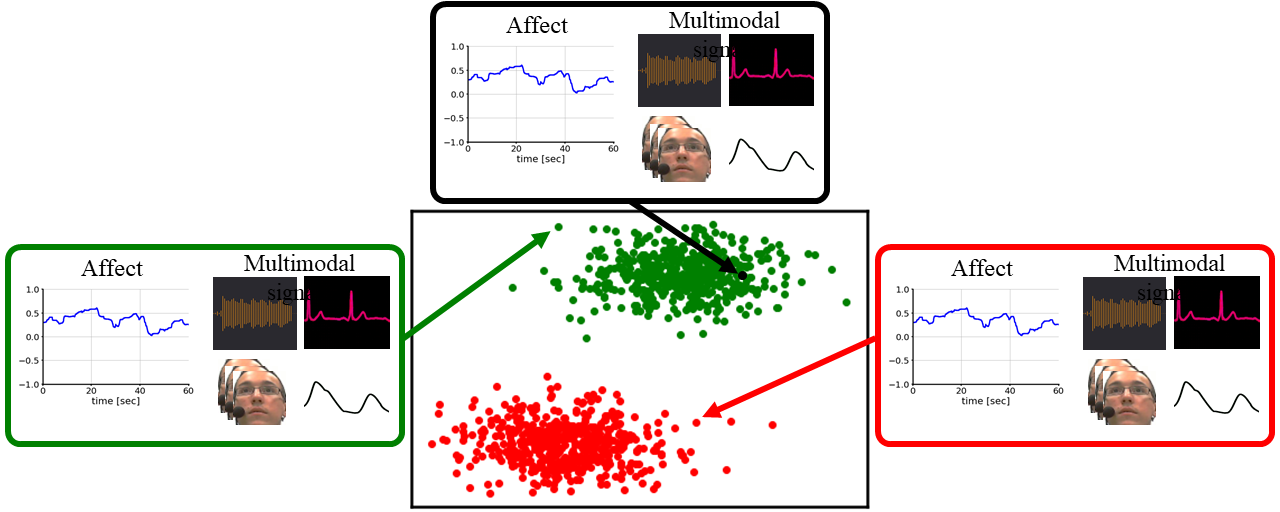}
\caption{
A high-level visualisation of the concept introduced. Supervised contrastive learning operates by infusing affect information within the representation, by pairing positive embeddings and dissociating negative embeddings. We assume affect is embedded in the multimodal latent space and defines what distinguishes (contrasts) data. Positive (green) and negative (red) multimodal data is labelled with respect to an anchor affect (black). Similar and dissimilar affect patterns define, respectively, positive and negative pairs of the anchor. The resulting representation is general with respect to affect patterns in a corpus.
}
\label{fig:concept}
\end{teaserfigure}



\maketitle

\section{Introduction}

Contrastive learning is a recent machine learning paradigm which has been widely and successfully employed for learning general representations of data \cite{saeed2021contrastive,le2020contrastive}. As a self-supervised learning method, it aims to project data into a space in which different views of the same input have similar representations \cite{le2020contrastive}. Although contrastive learning is quite popular within the computer vision domain \cite{diba2021vi2clr,jaiswal2020survey}, such methods have found applications in affective computing only recently for learning subject-invariant representations \cite{shen2022contrastive}. This paper builds on the hypothesis that affect information is an inherent property of the manifestations of affect and thus can be fused and learned in a contrastive learning manner, i.e. learning general affect-infused representations (see Fig. \ref{fig:concept}). In particular, we reframe the way multimodal affect modeling is viewed traditionally (end-to-end) and utilise affect annotations to build contrastive labels which are then used to train multimodal affect models.

To test our hypothesis that contrastive learning can learn general representations of affect and yield effective multimodal affect models, we build upon the Supervised Contrastive Learning (SCL) framework \cite{khosla2020supervised} for representation learning. In particular, we introduce and investigate three different approaches for building contrastive labels that rely on affect information (i.e. annotations). 
We evaluate the proposed methods against end-to-end classification using the extracted features of all modalities and corresponding arousal annotations of the RECOLA database \cite{ringeval2013introducing}. The approach investigated in this initial study feeds features captured within a time window into a neural network model trained via SCL. A probe model \cite{belinkov2021probing} is then employed to assess the quality of the learned representations by predicting high and low arousal states (i.e. arousal classification). Results indicate that SCL yields arousal models that perform \textit{significantly} better than end-to-end classification. This suggests that (a) contrastive learning is beneficial for downstream multimodal affect modelling tasks and (b) we can indeed fuse and blend affect information into the learnt representations through contrastive learning.

This paper is novel in several ways. First, to the best of our knowledge, this is the first time that the SCL framework has been employed in the context of affect modeling for learning affect-infused multimodal representations. Second, the generalization capacity of the representations learned via SCL is tested through three methods that utilise affect as a contrastive learning training signal. In particular, representations are trained on \textit{affect classes}, \textit{affect changes} and \textit{affect trends}. Finally, the proposed approach is compared against end-to-end classification across dissimilar modalities of the RECOLA database for predicting arousal, thereby extending the relatively sparse volume of work on the intersection of contrastive learning and multimodal affective computing.

\section{Related Work}\label{sec:related}

This section surveys related work on affect modelling using multiple and dissimilar user modalities, and moves on to review literature on the intersection of affect modelling and contrastive learning.

\subsection{Multimodal Affect Modeling}\label{sec:related_affect}

As emotions can be manifested through various user modalities including physiology, speech (audio) and facial expression (video) \cite{picard2000affective,calvo2015oxford} it is no surprise that aforementioned modalities of user data have been used extensively for modeling affect. When it comes to affect modeling based solely on visual information, the dominant approach for years has been to leverage domain knowledge and consequently to manually author high-level hand-crafted visual features or to employ classic pattern recognition techniques \cite{dahmane2011emotion,zheng2010emotion}. The advent of deep learning, however, allowed representation extraction to be automated and consequently revolutionized the field of pixel-based affect modeling. To the best of our knowledge, the study of Baveye et al. \cite{baveye2015deep} is the first to test the effectiveness of deep learning in the context of emotion prediction in videos. The authors concluded that although the size of the dataset might prevent the effectiveness of deep learning, using representations based on convolutional neural networks (CNNs) as features is promising for affective movie content analysis. Since then, deep learning has been used to extract relevant representations in many applications of affect modeling. Indicatively, Ng et al. \cite{ng2015deep} used a deep CNN pre-trained on the generic ImageNet dataset and leveraged transfer learning techniques to perform emotion recognition on small datasets. Makantasis et al. \cite{makantasis2019pixels} employed three dissimilar CNN architectures to predict player arousal from gameplay footage, showcasing that a mapping between gameplay video streams and the player's arousal exists. 

The field of emotion recognition via audio advanced significantly by leveraging domain knowledge, hand-crafted audio features and classic pattern recognition methods \cite{el2011survey,schuller2003hidden}. Nevertheless, deep learning algorithms quickly became a common practice in the field due to their high predictive capacity \cite{lieskovska2021review,abbaschian2021deep}. Indicatively, Huang et al. \cite{huang2014speech} learned candidate features via contractive CNNs and fed the learned features to a semi-CNN in order to extract affect-salient features. The experimental results showcased that the affect-salient features outperformed well-established features in speech emotion recognition. Kwon  \cite{kwon2021mlt} introduced a novel deep learning model for speech emotion recognition that utilizes a lightweight dilated CNN architecture and implemented the multi-learning trick approach. The model was evaluated on two benchmark datasets obtaining a high recognition accuracy. 

Emotion can be effectively captured by a subject's physiological response to a set of stimuli since physiological reactions (e.g. changes in brain activity and in somatic and visceral systems) are measurable manifestations of affect \cite{bradley2000measuring}. Early work in the field of psychophysiology focused on the use of affect models that map between hand-authored physiological features and affect \cite{mandryk2007fuzzy,holmgaard2015rank}. Martinez et al. \cite{martinez2013learning} introduced the first deep learning (CNN-based) approach for physiology-based affect modeling while Harper and Southern \cite{harper2020bayesian} presented an end-to-end deep learning model capable of classifying emotional valence from heartbeat time series along with a Bayesian framework for determining the uncertainty of the predictions. Giannakakis et al. \cite{giannakakis2019novel} employed a multi-kernel 1D CNN to compute complex feature maps performing multi-level modeling of the unique heart rate variability signature for stress identification.

Fusing more than one user modalities into an affect model (multimodal affect modeling) has been an active research field \cite{sebe2005multimodal,abdullah2021multimodal}; thus we will only focus on a few indicative studies. The work of Martinez et al. \cite{martinez2014deep} is arguably one of the first investigating the fusion of user modalities via a deep learning perspective. Recently, Makantasis et al. \cite{makantasis2021pixels} evaluated the capacity of deep learned representations to predict affect by relying only on audiovisual information of videos. Tzirakis et al. \cite{tzirakis2017end} employed a CNN and a deep residual network of 50 layers to extract features from speech and video modalities showcasing that deep learning can outperform traditional approaches in terms of concordance correlation coefficient. Guo et al \cite{guo2019multimodal} compared four combinations of eye images, eye movement and electroencephalograms and two fusion methods, demonstrating that different modalities provide complementary information for recognizing five emotions. Zhang et al. \cite{zhang2021deepvanet} employed a Convolutional LSTM to extract spatio-temporal facial features and a 1D CNN to extract bio-sensing features. Finally, the works in \cite{makantasis2021privileged,makantasis2021affranknet+} explored the concept of privileged information for fusing information from multiple modalities into unimodal affect prediction models.

In contrast to all aforementioned studies, in this paper we derive high-level representations that are relevant for the affect modelling task at hand through contrastive learning. The predictive capacity of these representations is tested in the task of arousal classification across the different modalities of the RECOLA database. Unlike previous studies which used either pre-trained models or models trained specifically (end-to-end) for each task, we test an intermediate solution which account for the particularities of the dataset but produces representations of affect that can be re-used for different downstream tasks.
 

\subsection{Contrastive Learning for Affect Modeling}\label{sec:related_contrastivelearning}

While research at the intersection of affective computing and contrastive learning (CL) algorithms has been active over the last few years, the literature is still relatively sparse. CL framework which utilizes a spatiotemporal augmentation scheme for facial expression recognition in videos. The same authors exploited facial images captured simultaneously from different angles and developed a two-step training process to address the Multi-view Facial Expression Recognition problem \cite{roy2021self}.
Li et al. \cite{li2021contrastive} investigated the impact of unsupervised representation learning on unlabeled datasets for speech emotion recognition and demonstrated that the contrastive predictive coding method can learn salient representations from unlabeled datasets that achieve state-of-the-art performance for the activation, valence and dominance primitives. Mai et al. \cite{mai2021hybrid} introduced a novel hybrid contrastive learning of tri-modal representation framework, using intra-/inter-modal and semi-contrastive learning to allow the model to explore cross-modal interactions and preserve inter-class relationships that reduce the modality gap. Finally, Yin et al. \cite{yin2021contrastive} proposed a two-step framework based on contrastive learning in order to address the cross-corpus emotion recognition problem. The authors demonstrate that utilizing contrastive learning to pre-train the encoder in a specific domain can produce representations that can be finetuned in a similar domain and achieve superior performance in the new domain.

In contrast to the aforementioned studies, this work contributes to the literature by introducing three different strategies for constructing the supervision signal used in SCL. All above-mentioned studies yield representations using solely information of the input space of the affect model. In this study, instead, we employ SCL to derive representations using affect-based contrastive labels. In other words, we fuse and blend affect information on the input space (i.e., subject measurements such as visual, audio and physiology data used in affect modelling) of learnt representations. Towards this direction, we exploit different statistical properties of continuous subjectively-defined affect annotations. Moreover, we investigate the impact of the different strategies on the expressiveness of the obtained representations and provide useful insights regarding their capacity to predict affective states. Results verify that employing contrastive learning representations boosts the performance of affect models for both uni- and multimodal affect measurements.

\section{Methodology}\label{sec:methodology}

This section describes the main elements of the algorithms examined in this paper. We start by presenting the representation learning components and move on with the methods employed for constructing the affect-based supervision signal for contrastive learning, i.e., contrastive labels, by selecting positive and negative samples based on continuous arousal annotation traces. The code for this paper is available on GitHub\footnote{\url{https://github.com/kpinitas/Supervised-Contrastive-Learning-for-Affect-Modeling}}.

\begin{figure*}[!tb]
\centering
\includegraphics[width=0.8\linewidth]{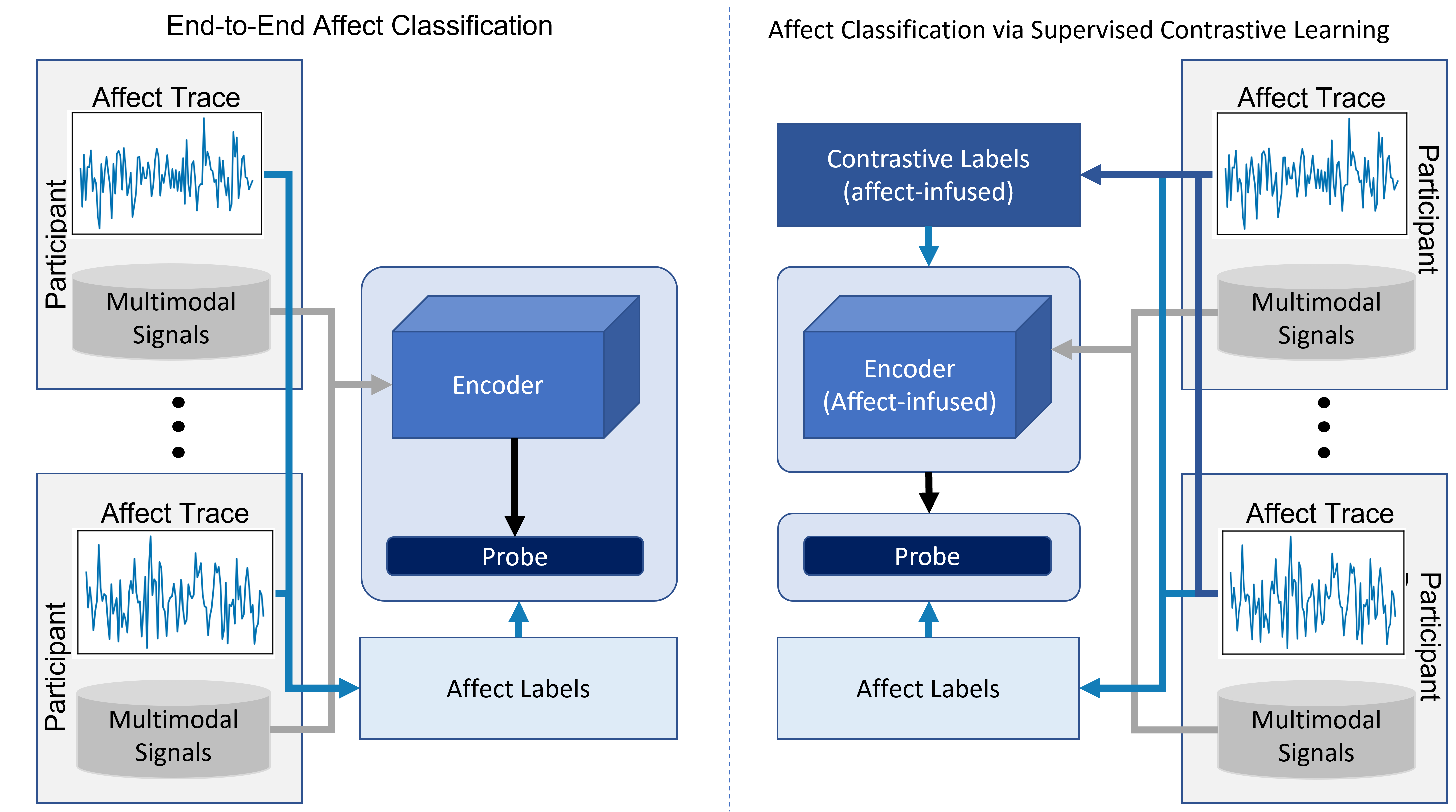}
\caption{
Illustration of the training methods employed: the end-to-end classification baseline (left) and the SCL method (right) which first derives affect-infused labels for contrastive learning, and then trains the probe model based on the affect representations of the pre-trained encoder. 
In both learning paradigms, affect labels are derived from participants' annotations.
}
\label{fig:architecture}
\end{figure*}

\subsection{Representation Components}\label{sec:methodology_models}

In this section we first present the main components used in representation learning, namely encoders and probes as depicted in Fig. \ref{fig:architecture}. Then, we present a baseline architecture of an end-to-end classifier that we compare against our SCL methodology for assessing the effectiveness of the obtained affect infused representations.

\subsubsection{Encoder}\label{sec:methodology_models_encoder}

An encoder model $E$ can be characterized by any neural network architecture that projects high dimensional data into a latent space of lower dimension, producing high-level representations of the input data. Hence, after training, $E$ is an efficient coding function that reduces the dimensionality of the data while maintaining essential information about the input space. In this work, we hypothesize that affect information is a manifested and embedded property of the input space and consequently, it can be merged with the latent space via contrastive learning to yield more robust representations. We thus derive contrastive labels of affect and subsequently train the encoder using the $L_{SCL}$ supervised contrastive loss function \cite{khosla2020supervised}:
\begin{equation}
L_{SCL} = \sum_{s\in S} \frac{-1}{|P_s|}\sum_{p \in P_s}\log\frac{\exp(r_s\cdot r_p/\tau)}{\sum_{a\in A_s}\exp(r_s\cdot r_a/\tau)}
\label{eq:loss_labels}
\end{equation}
\noindent where $S$ is a set that includes all samples and $P_s$ is the set that includes only the samples that are assigned to the same class as $s$ (positive sample set). In addition, $A_s$ is a set that contains any element of set $S$ besides element $s$. With $r_s$, $r_p$ and $r_a$ we denote the hidden representations of the model for the samples $s$, $p$ and $a$, respectively. Finally, $\tau$ stands for a non-negative temperature hyperparameter.   

\subsubsection{Probe}\label{sec:methodology_models_probe}

Probe architectures are mainly used to evaluate the quality of representations extracted from a pre-trained encoder $E$. In particular, given a known property (e.g. object categories in object recognition problems) of the input data, a probe can be trained to determine whether or not this property has been transferred to the latent space. Although there is no limitation in the number of hidden layers, a probe architecture usually consists of a single layer with softmax activation function, i.e. a \emph{linear probe} \cite{belinkov2021probing}.
  
\subsubsection{Baseline}\label{sec:methodology_models_baseline}

The baseline architecture used in this work realizes end-to-end classification (without the representation learning carried out via contrastive labels discussed in Section \ref{sec:methodology_contrastive}) and is labeled $E_b$. The model for $E_b$ consists of a randomly initialized encoder $E$ followed by a randomly initialized probe architecture. Both the encoder and the probe are trained (simultaneously) to map high-level handcrafted multimodal features to affect labels.

\subsection{Affect-Infused Contrastive Labels}\label{sec:methodology_contrastive}

The core contribution of this paper is the use of supervised contrastive learning algorithms for yielding general affect-infused representations that would be beneficial to any downstream affect modelling task. Specifically, we use SCL to pre-train the encoder architecture (see Section \ref{sec:methodology_models_encoder}) using affect-based contrastive labels to fuse affect information into representation learning. Then, we freeze the encoder's weights and train only the linear probe architecture (see Section \ref{sec:methodology_models_probe}) according to the downstream task. This training process is depicted in Fig. \ref{fig:architecture}.

Arguably one of the most crucial steps in contrastive learning is the choice of positive and negative samples, which highly affects the quality of the CL supervision signal. The selection process is relatively straightforward when the annotations can be defined objectively (e.g. class labels) since a positive pair usually consists of two samples of the same class and a negative pair consists of samples that belong to different classes \cite{khosla2020supervised}. However, when it comes to subjective annotations such as arousal signals, there is no clear pair selection strategy due to their subjective nature and data noise caused by the inherent bias of human annotators \cite{yannakakis2018ordinal}. 

\newcommand{\resw}{0.5\linewidth}
\newcommand{\resws}{0.37\linewidth}

This study explores three dissimilar affect measurements that can be calculated from any continuous annotation trace: one \emph{absolute} measure regarding the subject's current emotional state per se and two \emph{relative} measures regarding how the continuous emotional state fluctuates during a predefined time window \cite{yannakakis2018ordinal,camilleri2017generalmodels}.
We calculate the \textit{affect state} ($g_a$) as the mean of the affect trace captured within a time window (Eq. \ref{eq:ga}) \cite{makantasis2021pixels,melhart2022again}, the \textit{affect change} score ($c_a$) as the average of the absolute differences between consecutive annotation values (Eq. \ref{eq:ca}) and the \textit{affect trend} score ($t_a$) as the average of the differences between consecutive annotation values (Eq. \ref{eq:ta}):

\begin{align}
  g_a &= \frac{1}{w}\sum_{i=0}^{w} v_i
 	\label{eq:ga}\\
 	c_a &= \frac{1}{w}\sum_{i=1}^{w} |v_i-v_{i-1}|
 \label{eq:ca}\\
 	t_a &= \frac{1}{w}\sum_{i=1}^{w} \left( v_i-v_{i-1} \right)
 	\label{eq:ta} 
\end{align}
\noindent where $w$ is the window size considered and $v_i$ is the $i$-th annotation value of the time window. 

Based on the aforementioned measures, we explore three different positive/negative sample selection strategies that we detail below. It should be noted that, for all the proposed contrastive labelling strategies, we train the SCL models using the same loss function: i.e. the supervised contrastive loss function $L_{SCL}$ of Eq.~\eqref{eq:loss_labels}.

\subsubsection{Contrasting Affect: High vs. Low \label{sec:methodology_contrastive_hl}}

Intuitively, contrastive labels can be constructed by matching windows that have similar affect states as positive pairs and those with dissimilar affect states as negative pairs. To define affect state similarity, we binarize affect states $g_a$'s as ``high'' and ``low'', and consider windows with same (different) states as similar (dissimilar). The binarization criterion is based on the median ground truth value of the entire set of affect annotation traces ($\tilde{g}_a$) and on a threshold $\epsilon$. Specifically, a time window $i$ is labeled as ``high'' when $g_{a_i}>\tilde{g}_a+\epsilon$ and as ``low'' when $g_{a_i}<\tilde{g}_a-\epsilon$. It should be noted that the threshold $\epsilon$---as e.g. employed in \cite{makantasis2019pixels}---is used to eliminate windows with ambiguous affect annotation values close to the median which may deteriorate the stability of the SCL models and thus the effectiveness of the learned representations. The resulting preprocessed dataset that does not include the ambiguous affect values forms the basis for all three labeling strategies.

\subsubsection{Contrasting Affect: Change vs. Unchanged \label{sec:methodology_contrastive_change}}

While the high-low pairing strategy of Section \ref{sec:methodology_contrastive_hl} uses an absolute measure of affect, a similar binarization procedure can be performed based on \emph{affect change} (see Eq. \ref{eq:ca}) which is a relative measure. We can label a time window $i$ as ``change'' when $c_{a_i}>\tilde{c}_{a}$ and as ``unchanged'' (i.e. no change) when $c_{a_i}\leq\tilde{c}_{a}$. Once again, we set $\tilde{c}_{a}$ to be the median affect change value of the entire set of affect change traces. By selecting the median $\tilde{c}_{a}$ to binarize the data, we end up with a balanced dataset. As in Section \ref{sec:methodology_contrastive_hl}, we use these labels to pair windows $i$ and $j$ as positive when they both feature affect change or both feature no change, and as negative when one window features affect change and the other one does not.

\subsubsection{Contrasting Affect: Uptrend vs. Downtrend}
\label{sec:methodology_contrastive_trend}

Inspired by \cite{yannakakis2018ordinal} and similarly to the pairing strategy of Section \ref{sec:methodology_contrastive_change}, this contrastive labelling strategy uses a relative measure of affect, that of \emph{affect trend}. Hence the $i$-th time window is assigned to the ``uptrend'' class when $t_{a_i}>\tilde{t}_{a}$ otherwise it is assigned to the ``downtrend'' class. We set $\tilde{t}_{a}$ to be the median affect trend  value of the entire set of affect trend traces. Once again, we use the labels to define positive and negative samples based on the class that they belong to: a class match and a class mismatch define positive and negative pairs, respectively. 

The main difference between the first and the other two contrastive labelling strategies is that the former is direct as the ``high'' and ``low'' values are derived from the actual magnitude of the affect annotation trace (see Eq. \ref{eq:ga}). The other two strategies, instead, are indirect since both the ``change'' and the ``trend'' are higher order traces that express the average absolute rate of change (Eq. \ref{eq:ca}) and bent (Eq. \ref{eq:ta}) of the annotation trace, respectively. The binarization criterion for all three strategies, however, considers all the annotation traces of the affect corpus.

\section{The RECOLA Database}\label{sec:recola}

The methodology proposed in this paper is tested on a challenging dataset of online dyadic interactions between 23 participants that includes recordings and emotion annotations which are part of the RECOLA Database \cite{ringeval2013introducing}. RECOLA is a multimodal dataset that consists of audio, visual, and physiological recordings such as electrodermal activity (EDA), and electrocardiography (ECG). Aiming to strike a balance between the quantity and the quality of the features, RECOLA provides 40 handcrafted features for the video information, 116 for physiological signals and 130 for audio information. The video features correspond to action units, head pose estimation, and texture and optical flow features. The audio features are voicing related descriptors similar to those used in the COMPARE challenge \cite{schuller2013interspeech,schuller2014interspeech}. The physiological features extracted from EDA are related to the variability and zero-crossing of the signal as well as the spectral entropy and spectral coefficients. The ECG features instead include the spectral entropy and mean frequency of the signal. 

The collected data is annotated by six assistants (i.e. annotators) in terms of arousal and valence. The annotations are continuous, bounded in the range of $[-1,1]$, and provided at 25Hz. Apart from the features extracted from ECG, EDA signals, and raw footage information, the creators of the dataset also provide the videos from which the audiovisual features have been extracted. In this initial study, however, we do not make use of video frames (pixel information). It should be noted that amongst the 34 participants who gave their consent to share their data outside of the consortium, only 23 video recordings are publicly available. RECOLA has been used for audiovisual emotion recognition challenges in which the remaining 11 participants serve as an evaluation set, and thus their data is not publicly available.
 
\subsection{Data Preprocessing}\label{sec:recola_data}

The RECOLA database provides both arousal and valence annotations. This dataset has been a benchmark for the AVEC challenge for several years. Consequently, the Hall of Fame\footnote{Hall of Fame models: \url{https://diuf.unifr.ch/main/diva/recola/news.html}}  results can provide valuable insights into the quality of the features in predicting affect. In particular, the above results show that arousal is better captured in the provided audio, visual, and physiology features obtained by concatenating ECG and EDA features \cite{ringeval2015prediction} \cite{zhang2018dynamic} \cite{valstar2016avec}. For this reason, in this initial study, we focus on the comparison among the introduced SCL variants (regarding arousal) in deriving robust contrastive learning representations for arousal modeling. In particular, we aim to derive informative representations that capture the temporal information encoded in the provided features and use them to predict arousal states.

Towards this direction, we split each participant's session (features) into overlapping time windows using a sliding step of 400ms and window lengths of 1, 2, 3 and 4 seconds. The sliding step and window length are hyperparameters that affect the size of the dataset and the information contained in each window, respectively. It should be noted that the features and the arousal annotations are already synchronized, and we do not need to account for the reaction time between stimulus and emotional response. After splitting each session into time windows, each window consists of a sequence of  feature vectors. To reduce the computational load, we compute the average value for each feature inside the time window representing this way each time window by a single feature vector. Moreover, in this way the dimension of the feature vector is not dependant on the windows' length.  Table \ref{tab:number_of_samples} presents the number of samples per window length.

When it comes to affect annotation, we use the median annotation values per time window in order to mitigate inter-annotator disagreement \cite{grewe2007emotions}. The arousal state score $g_a$ is computed based on Eq.~\eqref{eq:ga} using the median arousal trace while $c_a$ and $t_a$ are computed according to Eq.~\eqref{eq:ca} and \eqref{eq:ta} where the consecutive value differences correspond to arousal value differences. The $g_a$ score is bound within $[-1, 1]$ as it captures the original
scale of arousal annotation, and measures the general arousal
level within the boundaries of the time window in question. The $c_a$
score is zero when the annotation value remains constant throughout the time window (i.e. no reported change in arousal) and is high when the annotation changes drastically (regardless of whether it increases, decreases, or both). Finally, $t_a$ is high when the annotation increases throughout the time window duration and low when the arousal score decreases. Note that unlike $g_a$, both $c_a$ and $t_a$ are unbounded, although in practice their values tend to be small.

\begin{table}[t]
\caption{Number of samples obtained by applying different time window lengths }\label{tab:number_of_samples}
\begin{center}
\begin{tabular}{|c|c|c|c|c|}

\hline
window length &\textbf{1 sec }&\textbf{2 sec}&\textbf{3 sec }&\textbf{4 sec} \\
\hline
sample size & 10235 &10124&9963&9811\\
\hline
\end{tabular}
\end{center}
\end{table}

\section{Results}\label{sec:results}

This section first outlines the experimental protocol we use for the evaluation of the methods introduced in this paper and then presents the key experimental results obtained.

\subsection{Experimental Protocol}

In this initial study, we test the proposed methods on the downstream task of arousal classification: the model has to learn to classify features within a time window as low or high arousal state. The encoder used in this work ($E$) is a simple ANN model with one sigmoid activated dense layer of $30$ trainable neurons. The output of the dense layer, given a time window, corresponds to the high-level representation describing this specific window. The probe architecture is a dense layer of two neurons activated using the softmax function. Finally, our baseline model $E_b$ is an encoder $E$ followed by a probe. Each model in this paper is trained using the Adam optimizer with learning rate of 0.001 
and batch size of 256. We set the temperature parameter $\tau$ in Eq.~\eqref{eq:loss_labels} to $\tau=0.1$. 
 
To generate the corresponding class labels (``high'' vs. ``low'' arousal) we follow the processes described in Section \ref{sec:methodology_contrastive_hl}. Moreover, the arousal change and arousal trend contrastive labels are generated only for the training samples of the downstream task to guarantee that the same training and test data are used for all models promoting a fair comparison among the presented algorithms. Furthermore, to evaluate the performance of each method, we use a five-fold cross-validation strategy for splitting the data into training and test sets ensuring that data in each set belong to different participants and thus the training and test sets are non-overlapping. The models are trained based on a convergence protocol that (early) stops the training process after 10 epochs without a training loss improvement and returns the model. The selected class splitting criterion (median of the arousal trace) and threshold $\epsilon=0.1$ ensure that both the training and test sets are balanced and consequently, the performance of the models is evaluated in terms of accuracy score. We present experiments for time window lengths of 1, 2, 3 and 4 seconds.

\begin{figure}[!tb]
\subfloat[1 second]{\includegraphics[width=\resw]{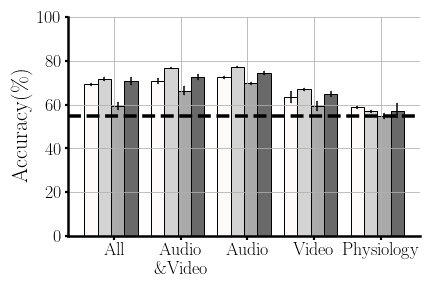}\label{fig:4a}}
\subfloat[2 seconds]{\includegraphics[width=\resw]{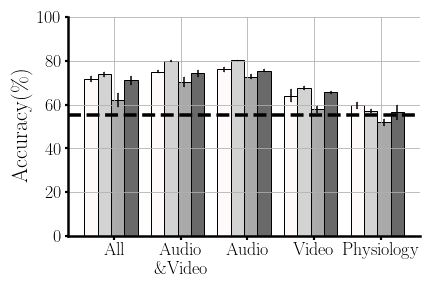}\label{fig:4b}}
\\
\subfloat[3 seconds]{\includegraphics[width=\resw]{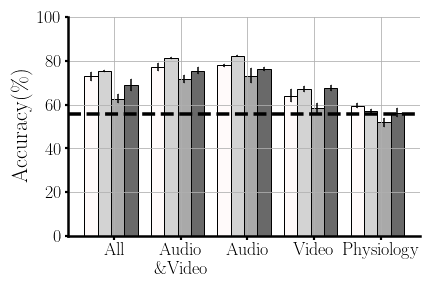}\label{fig:4c}}
\subfloat[4 seconds]{\includegraphics[width=\resw]{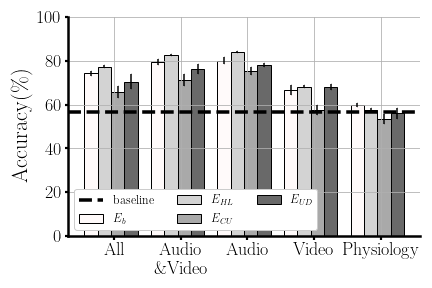}\label{fig:4d}}

\caption{
Average 5-fold validation accuracy scores ($\%$) for high-low arousal classification as a downstream task. Values are averaged across 10 independent runs; 95\% confidence intervals are displayed as error bars.
}
\label{fig:results}
\end{figure}

\subsection{Contrastive Learning For Arousal Modeling} \label{sec:results_hl}

We wish to investigate how representations learned via contrastive learning perform for a downstream task classifying between high and low arousal states. We thus train three encoders via SCL as per our three contrastive labeling strategies: high-low arousal state ($E_{HL}$), arousal change-unchanged ($E_{CU}$), and uptrend-downtrend ($E_{UD}$). 
We train a probe model for each of the encoders as mentioned in Section \ref{sec:methodology}. The baseline model, $E_b$, performs end-to-end arousal classification. An additional baseline always chooses the most frequent class in the training set (dotted line in Fig. \ref{fig:results}).

We explore five modality configurations of the RECOLA dataset: the single modalities of audio and video, the bimodal physiological signal containing ECG and EDA, the bimodal audio and video, and finally all modalities combined. Experiments across the four different modeling approaches, four time window lengths and five modality configurations are illustrated in Fig. \ref{fig:results}.
Results are based on 5-fold validation accuracy values, averaged across 10 independent runs. Statistical significance is established via a two-tailed Student's $t$-test with a significance threshold $p<0.05$.

Figure \ref{fig:results} indicates that $E_{HL}$ results in the best performing model. This is not surprising as the supervision signal in contrastive learning is the same as the downstream task. Specifically, the $E_{HL}$ model outperforms the baseline end-to-end classifier ($E_b$) yielding significantly higher accuracy scores across all window lengths and RECOLA modalities except physiology. Results also showcase that the $E_{UD}$ model performs on par and in some cases outperforms $E_b$, even though the contrastive learning supervision signal is different than the labels of the downstream task. Specifically, $E_{UD}$ outperforms $E_b$ significantly in 7 of 20 setups; for video input it outperforms $E_b$ on all time windows. However, $E_{UD}$ usually marks significantly lower accuracy scores than the $E_{HL}$ model (in 16 of 20 setups). The $E_{CU}$ model seems to perform poorly, reaching significantly lower accuracies than the $E_b$ model in all cases. This indicates that some distinctions (change-unchanged) for representation learning are too distant to the downstream task to be useful.

Comparing across RECOLA modalities, the models achieve the highest accuracy when arousal modeling relies on audio features across all time window lengths. High accuracy scores are also obtained by the models when the audio features are fused in the input space with video features (Audio \& Video) and with video and physiology (All) features. Although video features can yield robust arousal predictors, resulting models are inferior to the audio-based models. Finally, regardless of training method, all models underperform (at the same level as the baseline) when physiological signals are considered. It appears that arousal is not well manifested (or captured by) physiology in the RECOLA dataset.

Analysing some indicative key performance values obtained, it seems that $E_{HL}$ achieves the highest average accuracy (i.e. $83.9\%$) and highest best-fold accuracy (i.e. $87.6\%$) across ten independent runs when using audio features as input, for 4 second time windows. This corresponds to $4.7\%$ relative increase in accuracy compared to $E_b$. A similar pattern can be observed in the case of video-based arousal modeling where the highest average accuracy of $68.2\%$ (and  best-fold accuracy $73.8\%$) is achieved by $E_{HL}$ (a $2.3\%$ relative increase over $E_b$). It is worth noting that the performance of the models decreases for shorter time windows, but $E_{HL}$ retains the best accuracies in all cases except physiological data; indicatively with audio features $E_{HL}$ reaches an average accuracy of $77.2\%$ (a $6.5\%$ relative increase over $E_b$) on 1-second time windows. In stark contrast, when the models are trained solely on physiological features their performance is independent of the time window length as all models reach baseline performance. Physiological information seems irrelevant for the contrastive learning process as all SCL models perform worse than $E_b$.

\section{Discussion}\label{sec:discussion}

This work investigated the potential of contrastive learning for handling affect modeling tasks when the annotations are continuous and subjectively defined. We aimed to assess the quality of feature-based representations of affect learned via supervised contrastive learning by comparing the performance of the learned representations with the representations learned via traditional arousal classification (high-low arousal) across the different modalities of the RECOLA Database. The results indicate that the SCL encoders yield more reliable models of affect when the affect modeling task is treated as a classification task.

A worthwhile discussion is our choice of not applying any data augmentation \cite{shorten2019survey}. Although data augmentation is arguably a standard practice and consequently an integral part of the contrastive learning pipeline, we decided to omit this step since our input space in this initial study consists of hand-crafted features. Moreover, while data augmentation is prevalent in unsupervised contrastive learning, in our case we use affect labels for deriving positive pairs in a supervised fashion. It is worth exploring, however, whether additional data augmentation based on simple feature manipulation can augment the dataset and improve the models' performance.

In terms of future research, there are several directions that we can follow. An obvious next step towards generality is to test the efficiency of our model in predicting other affect dimensions such as valence or other core affective states such as happiness and fear. In addition, we plan to investigate the performance of our method in producing representations of affect when the affect modeling objective is treated as a regression or a preference learning task \cite{furnkranz2011preference,yannakakis2018ordinal}. Fine-tuning a large pre-trained model is a standard practice in contrastive learning when the model operates on the pixel-level of the image. Hence, another obvious next step for this work is to consider pixel-based information of image modalities and employ the proposed approach as a fine-tuning method for models that have already been trained on vast datasets (e.g. the ImageNet 1k dataset or more relevant face datasets). We did not use pixel-based representations and such pre-trained visual models in this initial study as we wanted to better investigate how the method performs with a simpler network and particularly compare it against an end-to-end training baseline. Finally, we note that the proposed methodology is general and thus applicable to any affective computing and classification task, as long as affect annotations exist and can be processed as labels.

\section{Conclusions}\label{sec:conclusions}

This paper introduced a representation (contrastive) learning method that views affect as a training signal and integral part of the learned representation. We presented three approaches for handling subjectively defined continuous annotations, realising supervised contrastive learning in the domain of affective computing. Our experiments showcased that it is possible to learn highly-performing general affect-infused representations from arousal annotations of the RECOLA dataset. Comparing our method against end-to-end classification---which is one of the standard learning paradigms for modeling affect---we observe that some of the proposed SCL methods lead to significant improvements in performance which, in turn, showcases that our approach yields more accurate and reliable models of affect. While this first demonstration of supervised contrastive learning for affect-based representation learning tasks already shows a high potential for affect modeling, additional experiments considering more tasks, datasets, and learning paradigms are needed to assess the capacity and efficacy of the proposed approach.

\begin{acks}
Kosmas Pinitas, Antonios Liapis and Georgios N. Yannakakis were supported by the European Union's H2020 research and innovation programme (Grant Agreement No. 951911). Konstantinos Makantasis was supported by the European Union's H2020 research and innovation programme (Grant Agreement No. 101003397).
\end{acks}

\bibliographystyle{ACM-Reference-Format}
\bibliography{contrastive_affect}


\end{document}